\documentclass[twoside,fleqn]{article}
\usepackage{espcrc2}
\usepackage{graphicx}

\newcommand{\ba}{\begin{array}{c}}
\newcommand{\ea}{\end{array}}

\newcommand{\be}{\begin{equation}}
\newcommand{\ee}{\end{equation}}
\newcommand{\chpt}{$\chi$PT}
\newcommand{\rcht}{R$\chi$T}

\newcommand{\mB}{\mathcal{B}}

\newcommand{\mM}{\mathcal{M}}
\newcommand{\mN}{\mathcal{N}}

\newcommand{\mT}{\mathcal{T}}
\newcommand{\Frac}[2]{\frac{\displaystyle #1}{\displaystyle #2}}
\newcommand{\cO}{{\cal O}}

\begin{document}

\title{
%
%
Relevance of final state interactions in
$\eta'\to\eta\pi\pi$ decays}

\author{J.J. Sanz-Cillero~\address{
  Istituto Nazionale di Fisica Nucleare INFN, Sezione di Bari,  \\
  Via Orabona 4, I-70126 Bary, Italy
 }
}

\begin{abstract}
A study of  the $\eta^\prime\to\eta\pi\pi$ Dalitz plot distribution
is presented in this talk.
The size of the branching ratio is properly understood
within $U(3)$ Chiral Perturbation Theory 
and Resonance Chiral Theory,  
in the framework of the $1/N_C$ expansion.
Nonetheless, unitarity effects   
must be   incorporated in order to achieve  an appropriate description
of the Dalitz slope parameters.
After taking the final state interactions
into account, our predictions become now
in agreement with the available experimental measurements,
although some clear differences show up with respect to
previous  theoretical estimates.

{\small
11.15.Pg,
%
12.39.Fe
%

Chiral Lagrangians, $1/N_C$ expansion, eta(958)
}

\end{abstract}

\maketitle


\vspace*{-8.5cm}\hspace*{15cm}
{ \small BARI$-$TH/638$-$10 }
\\
\vspace*{5.75cm}

\section{Introduction
}
\label{sec.intro}

In this talk
\footnote{
Contribution to the proceedings of
{\it Quark Confinement and the Hadron Spectrum IX},
August 30th -- September 3rd 2010, Madrid, Spain.
I would like to thank the organizers for their hospitality and
hard work in the organization of the conference.
This work was supported in part by
the Istituto Nazionale di Fisica Nucleare,
the Juan de la Cierva program,
the Ministerio de Ciencia e Innovaci´on under grant CICYT-FEDER-FPA2008-01430,
%
%
the EU Contract No. MRTN-CT-2006-035482, ``FLAVIAnet'',
the European Commission under the 7th Framework Programme through
%
%
the Spanish Consolider-Ingenio 2010 Programme CPAN (CSD2007-00042)
and
Generalitat de Catalunya under grant SGR2009-00894.
}
we present the results obtained for the
$\eta'\to\eta\pi\pi$ decay~\cite{etap-decay} within the framework of
chiral Lagrangians and the  $1/N_C$ expansion~\cite{NC}.
Since $G$-parity prevents intermediate vector mesons to contribute,
the scalars play a crucial role in  this decay,
specially the $f_0(600)$ (or $\sigma$) resonance even though the $a_0(980)$
is also present and, indeed, dominant for the branching ratio.
%

This decay is used here to test Chiral Perturbation Theory (\chpt ) \cite{ChPT} and its
extensions such as large--$N_C$  $U(3)$--\chpt\ \cite{U3-ChPT}  and
Resonance Chiral Theory (\rcht )~\cite{RChTa},
eventually providing predictions for some relevant  hadronic
parameters.
Recently, the GAMS-$4\pi$ and VES Collaborations have measured
the related Dalitz plot parameters which characterize the shape of the
decay~\cite{GAMSexp,VESexp},
complementing older results~\cite{Alde:1986nw}.   
New improved measurements are foreseen at KLOE-2, Crystal Ball,
Crystal Barrel and maybe WASA.

On the theory side, the $\eta^\prime\to\eta\pi\pi$ decays have been
studied within an effective chiral Lagrangian approach in which the
lowest lying scalar mesons are combined into a
nonet \cite{Fariborz:1999gr} and, more recently, within the framework
of $U(3)$ chiral effective field theory in combination with a 
coupled-channels approach \cite{Borasoy:2005du}.
Other older analyses based on chiral symmetry can be found
in~\cite{Cronin:1967jq}.

In the isospin limit considered all along the work
the charged and neutral decay amplitudes
coincide, although the neutral decay rate has an extra $1/2$ factor
due to phase-space symmetry.
The Dalitz plot distribution for the charged decay can be described
by the two kinematic  variables
$X=\frac{\sqrt{3}}{Q}(T_{\pi^+}-T_{\pi^-})$ and
$Y=\frac{m_\eta+2m_\pi}{m_\pi}\frac{T_\eta}{Q}-1$,
where $T_{\pi^\pm ,\eta}$ denote the kinetic energies of mesons in the
$\eta^\prime$ rest frame:
$T_\eta=\frac{(m_{\eta'}-m_\eta)^2-s}{2 m_{\eta'}}$,
$T_{\pi^+}=\frac{(m_{\eta'}-m_\pi)^2-u}{2 m_{\eta'}}$,
$T_{\pi^-}=\frac{(m_{\eta'}-m_\pi)^2-t}{2 m_{\eta'}}$
and
$Q=T_\eta+T_{\pi^+}+T_{\pi^-}=m_{\eta^\prime}-m_\eta-2m_\pi$.
The Mandelstam variables
$s\equiv (p_{\pi^+}+p_{\pi^-})^2$, $t\equiv (p_{\eta^\prime}+p_{\pi^-})^2$ and
$u\equiv (p_{\eta^\prime}+p_{\pi^+})^2$ have been employed here,
which obey the relation $s+t+u=m_{\eta'}^2+m_\eta^2+2m_\pi^2$.
The squared modulus of the decay amplitude
can be then expanded around the center of the Dalitz plot~\cite{etap-decay}:
\begin{eqnarray}
|\mM(X,Y)|^2&=&|\mN|^2
[1+(aY+dX^2)
\\
&& \hspace*{-0.5cm}
+(bY^2+\kappa_{21}X^2 Y+\kappa_{40} X^4)+\cdots
\label{eq.slope-par2}
\nonumber
\end{eqnarray}
Odd terms in $X$ are forbidden due to charge conjugation 
and the symmetry of the wave function.
%
%

\section{Large--$N_C$ \chpt}

Large-$N_C$ Chiral Perturbation Theory is an effective field theory where,
due to the large-$N_C$ limit ($N_C\rightarrow \infty$)~\cite{NC},
the singlet axial current is also conserved,
the chiral symmetry is enlarged to $U(3)$
and the $\eta'$ becomes the ninth Goldstone boson~\cite{U3-ChPT}.
A simultaneous expansion in powers
of momenta , quark masses and  $1/N_C$ is devised, such that
$p^2,\, m_{u,d,s},\, 1/N_C={\cal O}(\delta)$~\cite{U3-ChPT}.
%
%
At NLO, one finds the prediction~\cite{etap-decay}
%
\begin{eqnarray}
\label{amplLN}
\mM_{\eta'\to\eta\pi^+\pi^-} &=&\\
&&\hspace*{-2.85cm}
\,c_{qq}
\times  {1 \over  F^2}\Bigg[ \frac{m^2_{\pi}}{2}
- \frac{2 L_5 m^2_{\pi}}{F^2}\bigg( m^2_{\eta'}
+m^2_{\eta}+2 m^2_{\pi} \bigg)\, +
\nonumber\\
&&
\hspace*{-3.cm}
+ \frac{2(3L_2+ L_3)}{F^2}\bigg(s^2+t^2+u^2-(m^4_{\eta'}
+m^4_{\eta}+2 m^4_{\pi})  \bigg)
\nonumber\\
&&
\hspace*{-2.cm}
+ \frac{24 L_8 m^4_{\pi}}{F^2}
+ \frac{2}{3} \Lambda_2 m^2_{\pi}  \Bigg]\,
+ \, c_{sq} \times \frac{\sqrt{2} \Lambda_2 m^2_{\pi}}{3 F^2} \, ,
 \nonumber
\end{eqnarray}
with the NLO chiral low.energy constants (LECs)
$L_{2,3,5,8}$ and $\Lambda_2$, and
the $c_{qq}$ and $c_{qs}$ coefficients providing the
strange ($s$) and non-strange ($q$) component
in the external state $\eta - \eta'$~\cite{etap-decay}.
The employed values of $L_5$, $L_8$ and $\Lambda_2$~\cite{etap-decay}
are not actually crucial, since they are suppressed by powers of $m_\pi^2$.
The dominant term is given by  the combination $3L_2+L_3$,
which is fixed by means of the experimental branching ratio
$\mB_{\eta'\to\eta\pi^+\pi^-}=(43.2\pm 0.7)\%$~\cite{Nakamura:2010zzi}
and will be used to predict the Dalitz-plot parameters
which characterize the shape of the decay.

\section{Resonance Chiral Theory}

In the case of RChT at large $N_C$~\cite{RChTa}, one has
%
\begin{eqnarray}
\mM_{\eta'\to\eta\pi^+\pi^-} &=&
\\
&&\hspace*{-2.5cm}
c_{qq}\,\,\times\,\, \Frac{1}{  F_\pi^2} \,
\Bigg\{   \, \Frac{m_\pi^2}{2} \,\,\, +\,\, \,
\Frac{4 c_d c_m}{F_\pi^2}\, \Frac{m_\pi^4}{M_S^2}
\label{eq.RChT-amplitude}
\nonumber\\
&&\hspace*{-2.5cm}
\qquad   +
\Frac{1}{F_\pi^2}\Frac{\left[ c_d(t-m_\eta^2-m_\pi^2)+2 c_m  m_\pi^2\right]
\,}{M_{S}^2-t}
\nonumber\\
&&\hspace*{-0.25cm}
\times  [c_d (t-m_{\eta'}^2-m_\pi^2)+2 c_m m_\pi^2 ]
\nonumber\\
&&\hspace*{-2.5cm}
\qquad  +
\Frac{1}{F_\pi^2}  \Frac{\left[ c_d (u-m_\eta^2-m_\pi^2)+2c_m  m_\pi^2\right]
\,}{M_{S}^2-u}
\nonumber\\
&&\hspace*{-0.25cm}
\times  [c_d  (u-m_{\eta'}^2-m_\pi^2)+2c_m m_\pi^2 ]
\nonumber\\
&&\hspace*{-2.5cm}
\qquad \qquad
 +
\Frac{1}{F_\pi^2}
\,   \Frac{  \left[ c_d (s-m_\eta^2-m_{\eta'}^2)+2c_m  m_\pi^2\right]
\, }{M_{S}^2-s}
\nonumber\\
&&\times
 [ c_d (s-2 m_\pi^2)+2c_m m_\pi^2 ]  \Bigg\}
\, .  \nonumber
\end{eqnarray}
The largest contribution comes from the $c_d$ terms, which are proportional
to the external momenta. Everything else is proportional to $m_\pi^2$,
being   suppressed.
If one now performs the chiral expansion of the \rcht\
amplitude at low-energies
($s,t,u,m_P^2\ll M_S^2$),
the large--$N_C$ ChPT result~(\ref{amplLN})
is recovered up to contributions
subleading in $1/N_C$~\cite{etap-decay,RChTa}.

We used   
$M_S=980$~MeV for the scalar multiplet mass,
the resonance coupling $c_m$  from the high-energy
scalar form-factor constraint $c_m=\frac{F^2}{4 c_d}$~\cite{cdcm}
(not very relevant as it   always appears multiplied by a $m_\pi^2$ factor)
and the value $c_d=28.4$~MeV fixed through the experimental branching ratio,
$  \mB(\eta'\to\eta \pi^+\pi^-)=(43.2\pm 0.7)\%  $~\cite{Nakamura:2010zzi}.

In addition to the contribution from scalar resonances, one might
also consider the impact of  $J=2$ resonances.
Still, as the mass of the
lightest tensor multiplet  is roughly $M_{f_2}=1.2$~GeV,
one may just consider its
leading effect in $1/M_{f_2}^2$ rather than the whole
non-local resonance propagator structure.
Thus, it induces a contribution that has the form of the
$(3 L_2+L_3)$ term from  Eq.~(\ref{amplLN})~\cite{etap-decay},
where the tensor resonance contributions to the $\cO(p^4)$   LECs
($3 L_2^T+L_3^T= g_{f_2}^2/3 M_{f_2}^2=0.16\cdot 10^{-3}$)
were estimated in Ref.~\cite{Ecker-tensor}, after imposing
high energy constraints on the $\pi\pi$--scattering.

\begin{table*}
\begin{center}
\begin{tabular}{ccccccc}
\hline
  & $\rm \begin{array}{c}
U(3)-\chi PT~\cite{etap-decay}
\end{array}$
  & $\rm \begin{array}{c}
  R \chi T~\cite{etap-decay}
  \end{array}$
  &
  $\rm \begin{array}{c}     
  Th.~\cite{Borasoy:2005du}
  \end{array}$
  &
 $\rm \begin{array}{c}     
 Th.~\cite{Borasoy:2005du}
 \end{array}$
  &
  $\rm \begin{array}{c}     
  Exp.~\cite{GAMSexp}
  \end{array}$
  &
  $\rm \begin{array}{c}     
  Exp.~\cite{VESexp}
  \end{array}$
  \\
\hline
  a            $[Y]$           & $-0.098(48)^{\dagger}$
    &   - 0.098(48)$^\dagger$
  & -0.127(9) & -0.116(11) & -0.066(16)(3) & -0.127(16)(8)
  \\
  b          $[Y^2]$           & -0.0497(8)    &  - 0.0332(5)
  & -0.049(36) & -0.042(34)  & -0.063(28)(4)  &  -0.106(28)(14)
  \\
  d          $[X^2]$           & -0.092(8)    &  -0.0718(3)
  &   +0.011(21)   &  +0.010(19)  &  +0.018(78)(6)   &  -0.082(17)(8)
  \\
  $\kappa_{21}$    $[X^2Y]$    &  0.003(2)     &  -0.009(2)
  & --- & --- & --- & ---
  \\
  $\kappa_{40}$    $[X^4]$     & 0.0022(4)     &  0.0013(1)
  & --- & --- & --- & ---
  \\
\hline
\end{tabular}
\caption{{\small
The results for the $N/D$--unitarized $\chi$PT amplitude are given in
the second column~\cite{etap-decay}.
In the third column we consider the$N/D$--unitarization
of  \rcht , including the contribution from $J=2$
resonances~\cite{etap-decay}.
In both cases, the $a$ parameter was taken as input.
The fourth and fifth columns provide previous theoretical predictions
for, respectively,  the $\eta'\to\eta\pi^0\pi^0$ and
$\eta'\to\eta\pi^+\pi^-$ decays~\cite{Borasoy:2005du}.  The last two
columns contain the experimental measurements from
GAMS-$4\pi$~\cite{GAMSexp} and VES~\cite{VESexp}, respectively.
}}
\label{tab.results}
\end{center}
\end{table*}

\section{Unitarization}

The narrow width of the $a_0(980)$~\cite{Nakamura:2010zzi}
and the smallness of the $\eta\pi$ scattering-length~\cite{Kubis1}
seem to point out  the  little relevance of the rescattering
in this channel.
Thus, in the elastic region
(no other channel opens up in the $\eta'$--decay  phase-space),
one has  the approximate $s$--channel unitarity relation
for the $\pi\pi$ rescattering,
\begin{equation}
\mbox{Im}\mM_J(s) \,\,=\,\, \rho(s)\,\mT^0_J(s)^* \, \mM_J(s)\, ,
\label{eq.unitarity}
\end{equation}
where
the decay amplitude $\mM(s,t,u)$ has been   decomposed into partial waves
$\mM_J(s)$ in the $\pi\pi$ angle $\theta_\pi$~\cite{etap-decay},
$\rho(s)=\sqrt{1- 4m_\pi^2/s}$  and $\mT^{I=0}_J(s)$ is the isoscalar
$ \pi\pi$ partial-wave scattering amplitude.
The absorptive cuts in the $t$ and $u$ $\eta\pi$--channels
have been neglected in~(\ref{eq.unitarity}).

There are various options  for the reconstruction
of the unitarized amplitude as the optical theorem only refers to
the absorptive part of the amplitude.  In our opinion,
the $N/D$--method~\cite{ND} is the most reliable one, as it
also incorporates the real part of the logarithm that
arises in the two-propagator
Feynman integral $B_0(s,m_\pi^2,m_\pi^2)$ at one loop,
not only its imaginary part $\rho(s)/16\pi$:
\begin{eqnarray}
\mM(s,t,u)^{\rm N/D} &=&  \sum_J 32\pi\,  (2 J+1)\,  P_J(cos\theta_\pi)
\nonumber\\
&& \hspace*{-1.cm}
\times \, \Frac{M_J(s)^{\rm tree}}{1\, -\, 16\pi B_0(s ,m_\pi^2,m_\pi^2
)\, \mT^0_J(s)^{\rm tree}}\, ,
\nonumber\\
\end{eqnarray}
with the Legendre polynomials $P_J(x)$, the partial wave decomposition
of the previously computed tree-level amplitudes
$\mM_J(s)^{\rm tree}$ and $\mT_J^0(s)^{\rm tree}$,  and
$16\pi^2 B_0(s,m_\pi^2,m_\pi^2)= C-\rho(s)\ln\frac{\rho(s)+1}{\rho(s)-1} $.
%
%
Actually, the integral $B_0$   is ultraviolet divergent and has a local
indetermination $C$  (denoted through $a^{SL}(s_0)$ in the
$N/D$ analysis~\cite{ND}) which requires an extra renormalization condition.
We will fix it by means of the experimental range for the
Dalitz-parameter $a=-0.098(48)$~\cite{etap-decay,GAMSexp,VESexp}.


\section{conclusions and discussion}

In this talk we have presented some new results on the $\eta'\to\eta\pi\pi$
decay within the large--$N_C$ \chpt\ and  \rcht\
frameworks~\cite{etap-decay}.
In both of them, the order of magnitude of the experimental branching ratio
is conveniently understood.
%
%
Furthermore, we obtained successful predictions for
the Dalitz slope parameters based on the
$s$--channel unitarization of our tree-level amplitudes~\cite{etap-decay}.
These have been summarized in Table~\ref{tab.results}and
compared to other theoretical predictions~\cite{Borasoy:2005du}  and
experimental measurements~\cite{GAMSexp,VESexp}.
Preliminary results from BES-III seem to provide
much more precise determinations:
$a=-0.047(12)$, $b=-0.068(21)$ and $d=-0.073(13)$~\cite{preliminary-BESIII}.
This clearly would favour our determinations with respect to
previous theoretical studies which predict a very small or positive
Dalitz parameter $d$~\cite{Borasoy:2005du}.
%
%
Future experiments
will be able to discern what is the most convenient framework for the
study of this and other $\eta'$ decays.

\end{document}